\documentclass[aip,reprint]{revtex4-1}
\usepackage{hyperref}
\usepackage{graphicx}

\draft 

\begin{document}


\title{Generic Fe buffer layers for Fe-based superconductors: Epitaxial FeSe$_{1-x}$Te$_x$ thin films} 

\author{Kazumasa\,Iida}
\email[Electronical address:\,]{k.iida@ifw-dresden.de}
\author{Jens\,H\"{a}nisch}
\affiliation{Institute for Metallic Materials, IFW Dresden, D-01171 Dresden, Germany}
\author{Michael\,Schulze}
\author{Saicharan\,Aswartham}
\author{Sabine\,Wurmehl}
\author{Bernd\,B\"{u}chner}
\affiliation{Institute for Solid State Research, IFW Dresden, D-01171 Dresden, Germany}
\author{Ludwig\,Schultz}
\author{Bernhard\,Holzapfel}
\affiliation{Institute for Metallic Materials, IFW Dresden, D-01171 Dresden, Germany}

\date{\today}

\begin{abstract}
Biaxially textured FeSe$_{1-x}$Te$_x$ films have been realized on Fe-buffered MgO substrates by pulsed laser deposition. Similar to the Fe/BaFe$_2$As$_2$ bilayers, the crystalline quality of FeSe$_{1-x}$Te$_x$ films exhibit a sharp out-of-plane and in-plane texture less than 0.9\,$^\circ$. The Fe/FeSe$_{1-x}$Te$_x$ bilayers showed high superconducting transition temperatures of over 17\,K. The angular-dependent critical current densities exhibit peaks positioned at $H$\,$\perp$\,${c}$ similar to other pnictide thin films. The volume pinning force of FeSe$_{1-x}$Te$_x$ in this direction is very strong compared with that of Co-doped BaFe$_2$As$_2$, due to a good matching between the interlayer distance in the $c$ direction and the out-of-plane coherence length.  
\end{abstract}


\maketitle
Epitaxial iron chalcogenide superconducting thin films have been prepared on single crystalline oxides substrates by pulsed laser deposition (PLD) or molecular beam epitaxy (MBE) to investigate their intrinsic properties and to explore superconducting device applications.\cite{01,02,12,04,05,06} These superconductors are very sensitive to strain.\cite{05} Indeed, the superconducting transition temperature ($T_{\rm c}$) of FeSe$_{0.5}$Te$_{0.5}$ films can be tuned by strain and even higher $T_{\rm c}$ values in films than in bulk materials can be realized.\cite{03} In addition, the superconductivity is induced in FeTe films by the tensile stress albeit their bulk crystals are not 
superconducting at ambient pressure.\cite{24}

The biaxial strain is usually induced by a lattice mismatch between films and substrates. However, the correlation between $T_{\rm c}$ and lattice mismatch for FeSe$_{0.5}$Te$_{0.5}$ is controversial.\cite{07} A fundamental problem is the formation of an interfacial layer between FeSe$_{0.5}$Te$_{0.5}$ and oxide substrates,\cite{07} which compromises the epitaxial growth. Later the interfacial layer was identified as amorphous containing oxygen, which comes mainly from oxide substrates.\cite{08} Furthermore, oxygen diffusion into the films was detected by transmission electron microscope (TEM) analyses, which deteriorates the superconducting properties. In order to minimize the oxygen diffusion, Tsukada $et$ $al$. have proposed the implementation of CaF$_2$ substrates, resulting in good superconducting properties.\cite{09} Hence, ideal growth conditions of FeSe$_{1-x}$Te$_x$ thin films require non-oxide substrates under ultra-high vacuum (UHV) atmosphere.

Recently we have reported that epitaxial Co-doped BaFe$_2$As$_2$ (Ba-122) films can be realized on Fe-buffered single crystalline MgO substrates in UHV condition.\cite{10} The detailed TEM analyses revealed that the FeAs tetrahedron in the Ba-122 bonds coherently to body centered cubic Fe. This Fe/Ba-122 bilayer has a clean microstructure and is of excellent crystalline quality without grain boundaries (GBs).\cite{11} Since the FeAs or FeSe(Te) tetrahedron is a common structure for both Ba-122 and FeSe$_{1-x}$Te$_x$, the implementation of an Fe buffer should be also applicable for the epitaxial growth of FeSe$_{1-x}$Te$_x$. In this letter, we demonstrate the implementation of an Fe buffer layer to grow epitaxial FeSe$_{1-x}$Te$_x$ films and present their transport properties.

\begin{figure}
	\centering
		\includegraphics[width=\columnwidth]{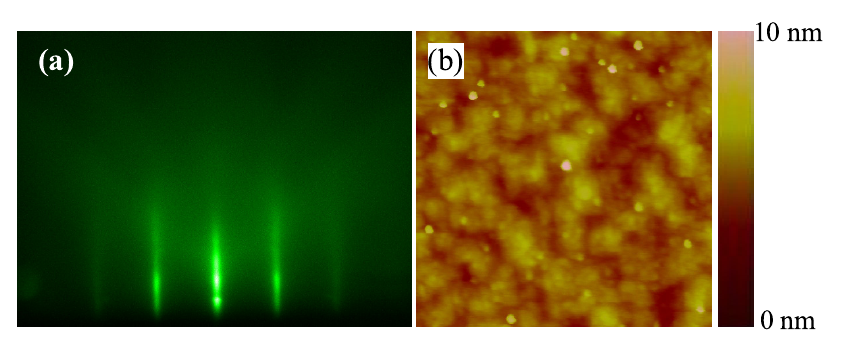}
		\caption{(Color online) (a) The RHEED image of FeSe$_{1-x}$Te$_x$ at room temperature shows only streaks, indicative of a flat surface. The incident electron beam is the MgO [110] azimuth. (b) The corresponding AFM image (1$\mu$m$\times$1$\mu$m) shows an $R_{\rm rms}$ value of 0.73\,nm.} 
\label{fig:figure1}
\end{figure}

The Fe/FeSe$_{1-x}$Te$_x$ bilayers were prepared by PLD (KrF excimer laser, $\lambda=248\,{\rm nm}$), which is almost identical deposition methods to Fe/Ba-122 except for the deposition temperature and the laser repetition rate of FeSe$_{1-x}$Te$_x$. The energy density of the laser on the target was 3-5\,J/cm$^2$ and the distance between target and substrate was approximately 6\,cm. Single crystalline MgO (100) substrates were heated to 1000\,$^\circ$C, held at this temperature for 30\,min, and subsequently cooled to room temperature for cleaning. After the Fe deposition at room temperature, the Fe-covered MgO was heated to 750\,$^\circ$C, held at this temperature for 20\,min and subsequently cooled to 450\,$^\circ$C, the optimum deposition temperature of FeSe$_{1-x}$Te$_x$ films with highest $T_{\rm c}$ value, similarly to the results in Ref.\cite{12}. Once the temperature was stabilized, the FeSe$_{1-x}$Te$_x$ layer was deposited at a laser frequency of 3\,Hz. The whole deposition process was conducted under UHV condition (base pressure of 10$^{-10}$\,mbar). The respective layer thickness of Fe and FeSe$_{1-x}$Te$_x$ were 20\,nm and 95\,nm confirmed by cross-sectional focused ion beam cuts at different sample areas as well as TEM.

The PLD target was prepared by a modified Bridgman technique yielding an Fe-Te-Se crystal with the nominal composition of Fe:Se:Te=1:0.5:0.5. For the growth, stoichiometric amounts of pre-purified metals were sealed in an evacuated quartz tube. The tube was placed in a horizontal tube furnace and heated up to 650\,$^\circ$C and kept at that temperature for 24\,h. The furnace was then heated to 950\,$^\circ$C and the temperature was kept constant for 48\,h. Finally, the furnace was cooled down with a rate of 5\,$^\circ$C/h to 770\,$^\circ$C, followed by furnace cooling. We yield crystals with dimensions up to cm-size. A bulk $T_{\rm c}$ of 13.6\,K was recorded by a superconducting quantum interface device (SQUID) magnetometer. Details of the single crystal preparation and their properties can be found in Ref. \cite{19}.

Each deposition step was monitored by reflection high-energy electron diffraction (RHEED). The RHEED image of the FeSe$_{1-x}$Te$_x$ acquired at room temperature, fig.\,\ref{fig:figure1}\,(a), shows only streaks similarly to the Fe buffer layer. This is indicative of a smooth surface of the FeSe$_{1-x}$Te$_x$ layer, which is consistent with the observation by atomic force microscope (AFM) presented in fig.\,\ref{fig:figure1}\,(b). A root mean square roughness ($R_{\rm rms}$) of 0.73\,nm was recorded. 

\begin{figure}
	\centering
		\includegraphics[width=\columnwidth]{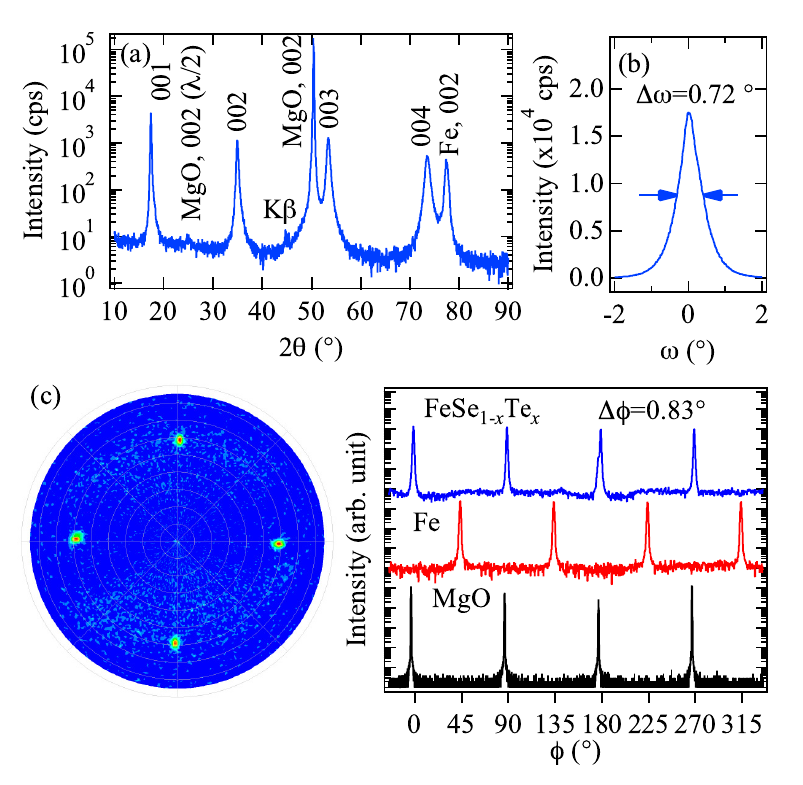}
		\caption{(Color online) (a) The $\theta\rm/2\theta$-\,scan of an FeSe$_{1-x}$Te$_x$ bilayer in Bragg-Brentano geometry using Co-K$\alpha$ radiation. Both layers of Fe and FeSe$_{1-x}$Te$_x$ were grown with $c$-axis texture. (b) Rocking curve of the 001 reflection. $\Delta\omega=0.72^{\circ}$ indicates good out-of-plane texture. (c) The 101 pole figure measurement and the corresponding $\phi$-scans of FeSe$_{1-x}$Te$_x$. The respective $\phi$-scans of the 110 Fe and the 220 MgO are also presented. The average $\Delta\phi$ of Fe and FeSe$_{1-x}$Te$_x$ are 1.05$^{\circ}$ and 0.83$^{\circ}$, respectively.} 
\label{fig:figure2}
\end{figure}

In order to check phase purity and texture quality of the films, detailed structural characterizations by x-ray diffraction were conducted as summarized in fig.\,\ref{fig:figure2}. The $\theta\rm/2\theta$\,-\,scans, fig.\,\ref{fig:figure2}\,(a), show only 00$l$ reflections of FeSe$_{1-x}$Te$_x$ together with the 002 reflection of Fe and MgO, indicating $c$-axis texture and phase purity. The $\omega$\,-\,scan for the 001 reflection of FeSe$_{1-x}$Te$_x$ in fig.\,\ref{fig:figure2}\,(b) shows a sharp full width at half maximum (FWHM, $\Delta\omega$) of 0.72$^{\circ}$. The 101 pole figure measurements and the corresponding $\phi$-scans of FeSe$_{1-x}$Te$_x$ reveal no satellite and additional reflections other than sharp and strong reflections at every 90$^{\circ}$, indicative of biaxial texture. Here, the epitaxial relation is identified as (001)[100]FeSe$_{1-x}$Te$_x$$\|$\\(001)[110]Fe$\|$(001)[100]MgO. The average $\Delta\phi$ of Fe and FeSe$_{1-x}$Te$_x$ are 1.05$^{\circ}$ and 0.83$^{\circ}$, respectively. From these results, the FeSe$_{1-x}$Te$_x$ layer is of good crystalline quality without GBs.

\begin{figure}
	\centering
		\includegraphics[width=\columnwidth]{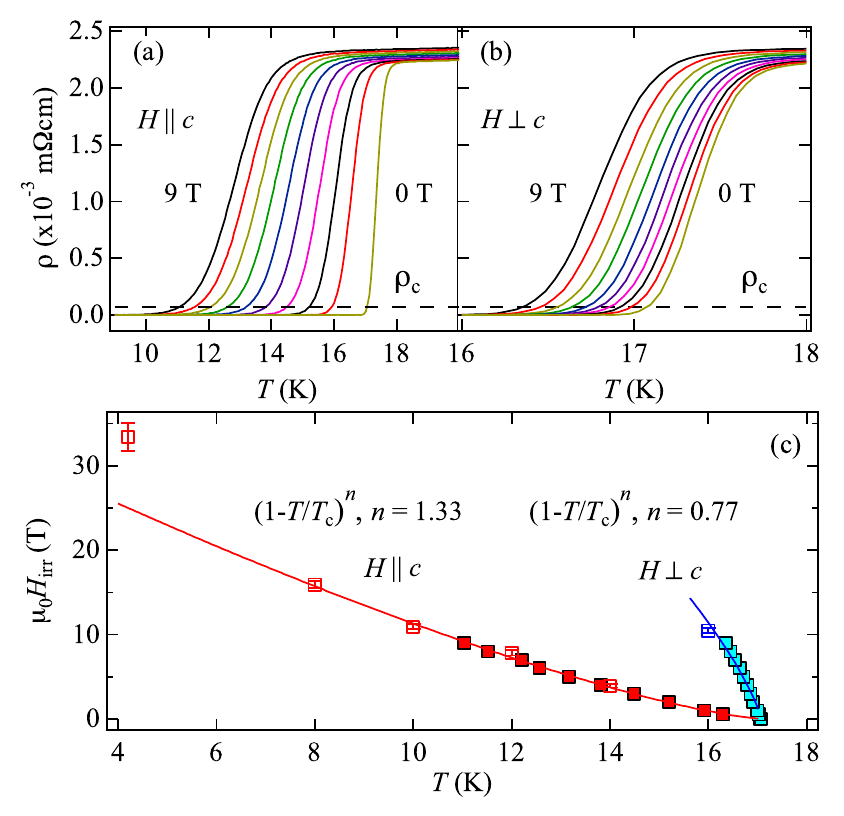}
		\caption{(Color online) (a) Resistivity traces for FeSe$_{1-x}$Te$_x$ film measured in several magnetic fields for $H$\,$\parallel$\,${c}$ and (b) $H$\,$\perp$\,${c}$. Field increment is 1\,T. Broken lines indicate the $\rho_{\rm c}$ for $\mu_0H_{\rm irr}$.(c) $\mu_0H_{\rm irr}$ for $H$\,$\parallel$\,${c}$ are always lower than that for $H$\,$\perp$\,${c}$. Open symbols are evaluated by the Kramer extrapolation.} 
\label{fig:figure3}
\end{figure}

In order to avoid any damage during ion beam etching and to ensure a low contact resistance, an Au layer was deposited on the film at room temperature by PLD. The Au-covered FeSe$_{1-x}$Te$_x$ film was then ion beam etched to form bridges with 0.5\,mm width and 1\,mm length for transport measurements, which were conducted in a Physical Property Measurement System (PPMS, Quantum Design) by a four-probe method. Here, a criterion of 1\,$\rm\mu Vcm^{-1}$ ($E_{\rm c}$) was used for evaluating $J_{\rm c}$.

The FeSe$_{1-x}$Te$_x$ film exhibits no sign of a resistance anomaly at around 200\,K, which has been frequently observed in FeSe$_{1-x}$Te$_x$ thin films. In addition, relatively small resistivity and a high residual resistivity ratio of 3.5 is recorded. This is a consequence of the current shunting effect since the resultant film was fully covered with Au. Therefore, the normal state behavior is masked by Au. Nevertheless, the onset $T_{\rm c}$ is recorded at 17.7\,K, which is higher than the bulk value. The higher $T_{\rm c}$ might be due to the strain effect as reported in Ref.\cite{03}. When magnetic fields are applied to the film, a clear shift of $T_{\rm c}$ to lower temperatures is observed for both directions, as exhibited in figs.\,\ref{fig:figure3}\,(a) and (b). In particular, this shift together with a broadening of the transition is more significant for $H$\,$\parallel$\,${c}$ than $H$\,$\perp$\,${c}$. Figure\,\ref{fig:figure3}\,(c) shows the temperature dependence of the irreversibility field ($H_{\rm irr}$), which is defined by the intersection between the resistivity traces and the resistivity criterion ($\rho_{\rm c}$) as shown in figs.\,\ref{fig:figure3}\,(a) and (b). In addition, $\mu_0H_{\rm irr}$ evaluated by the Kramer extrapolation from $J_{\rm c}-H$ characteristics presented later are also plotted.\cite{25} Here the $\rho_{\rm c}$ is defined as $E_{\rm c}$/$J_{\rm c,15}$, where the $J_{\rm c,15}=15\,{\rm Acm^{-2}}$ is a criterion for $H_{\rm irr}$ in $J_{\rm c}-H$ characteristics. In this definition, the $N$-value of the voltage-current characteristics ($V-I$, $V\propto I^N$) is 1 in the vicinity of $E_{\rm c}$. It is clear from fig.\,\ref{fig:figure3}\,(c) that $\mu_0H_{\rm irr}$ for $H$\,$\parallel$\,${c}$ is always lower than that for $H$\,$\perp$\,${c}$, indicating that the flux pinning is anisotropic. The $\mu_0H_{\rm irr}$ for $H$\,$\parallel$\,${c}$ shows a power law relation, $\mu_0H_{\rm irr}\sim(1-T/T_{\rm c})^n$, with exponent $n=1.33\pm0.02$ in the temperature range from 8\,K to 17\,K. At 4.2\,K, an increased value for $\mu_0H_{\rm irr}$ is observed presumably due to multi-band effects.\cite{16} On the other hand, an exponent of $n=(0.77\pm0.05)<1$ is deduced for $H$\,$\perp$\,${c}$, which is very similar to the upper critical field ($\mu_0H_{\rm c2}$) for $H$\,$\perp$\,${c}$ in layered compounds (i.e. $\mu_0H_{\rm c2} \propto (1-T/T_{\rm c})^{0.5}$).\cite{21}         

\begin{figure}
	\centering
			\includegraphics[width=\columnwidth]{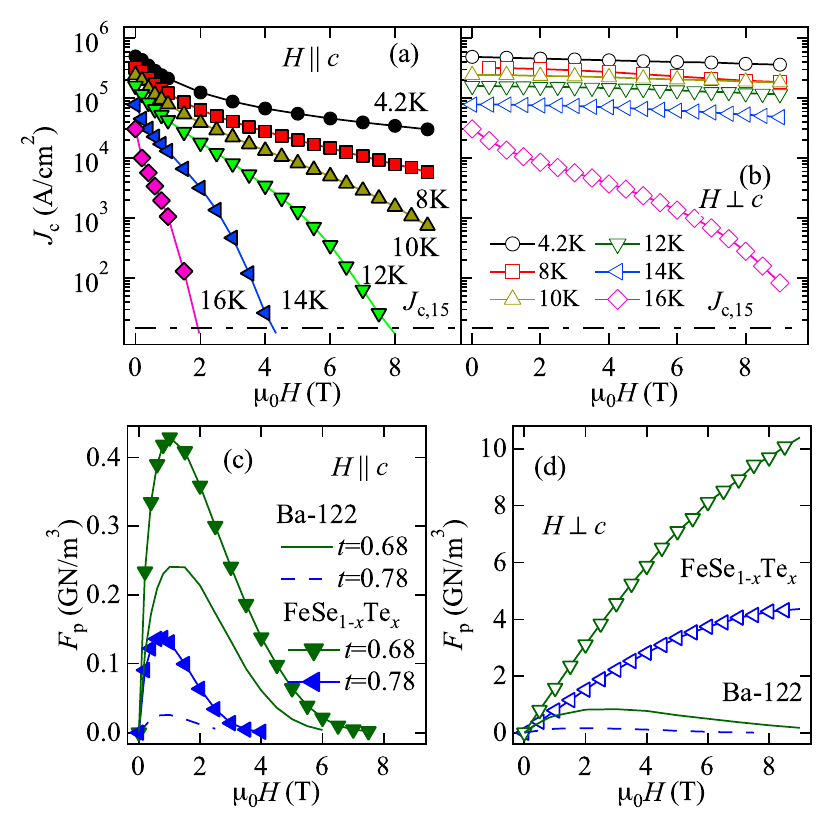}
		\caption{(Color online) (a) $J_{\rm c}-H$ characteristics measured in $H$\,$\parallel$\,${c}$ and (b) $H$\,$\perp$\,${c}$ at several temperatures. Chain lines indicate a $J_{\rm c}$ criteria of $15\,{\rm Acm^{-2}}$ for $\mu_0H_{\rm irr}$. (c) A comparison of the $F_{\rm p}$ between FeSe$_{1-x}$Te$_x$ and Co-doped Ba-122 ($T_{\rm c}=23.1$\,K) at reduced temperatures of $t=0.68$ and 0.78 for $H$\,$\parallel$\,${c}$ and (d) $H$\,$\perp$\,${c}$. The data of Co-doped Ba-122 plotted as lines are taken from the Ref.\cite{11}.} 
\label{fig:figure4}
\end{figure}

Shown in fig.\,\ref{fig:figure4}\,(a) are $J_{\rm c}-H$ characteristics measured for $H$\,$\parallel$\,${c}$ at several temperatures. $J_{\rm c}$ values are quickly reducing with increasing $H$, indicative of less flux pinning in this direction. It should be noted that $V-I$ curves for evaluating $J_{\rm c}$ always show a power-law relation, suggesting that current is limited by depinning of flux rather than GBs. On the other hand, the reduction of $J_{\rm c}$ with $H$ is very small for $H$\,$\perp$\,${c}$ below 16\,K (fig.\,\ref{fig:figure4}\,(b)). The Se(Te)-Se(Te) interlayer distance in the $c$ direction is almost identical to the out-of-plane coherence length at low temperatures ($\xi_{\rm c}$=0.35\,nm), resulting in strong pinning in this direction. Here the Se(Te)-Se(Te) interlayer distance is calculated as around 0.25\,nm using $c(1-2z)$ with the lattice parameter $c$ and the Se(Te) coordination $z$.\cite{05} For Co-doped Ba-122, the As-As interlayer distance in the $c$ direction is 0.3\,nm, which is far below $\xi_{\rm c}$ (1.2\,nm). As a result, $J_{\rm c}$ at $H$\,$\perp$\,${c}$ is decreased significantly with increasing $H$ compared with FeSe$_{1-x}$Te$_x$. The volume pinning force ($F_{\rm p}$) of FeSe$_{1-x}$Te$_x$ is larger than that of Co-doped Ba-122 for both major directions at reduced temperatures of $t=0.68$ and 0.78 (fig.\,\ref{fig:figure4}\,(c) and (d)). However, $F_{\rm p}$ of Co-doped Ba-122 for $H$\,$\parallel$\,${c}$ is larger than that of FeSe$_{1-x}$Te$_x$ at $t=0.48$ (not shown in this letter). The origin of this observation has to be investigated in detail by TEM in future.

\begin{figure}
	\centering
			\includegraphics[width=\columnwidth]{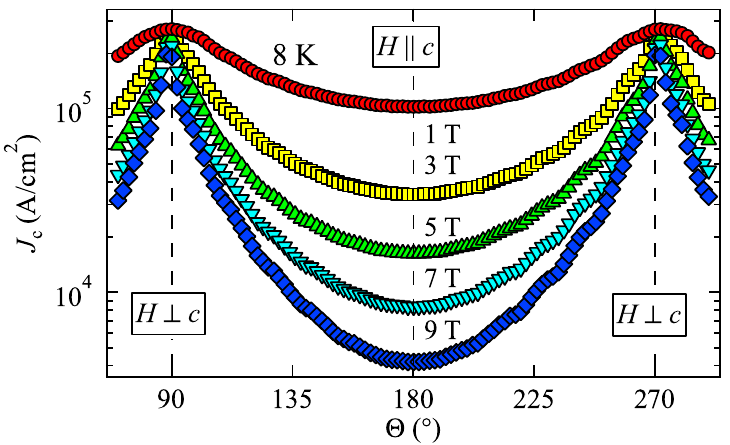}
		\caption{(Color online) $J_{\rm c}(\Theta)$ of the FeSe$_{1-x}$Te$_x$ film measured in several magnetic fields at 8\,K. No additional peaks except at $H$\,$\perp$\,${c}$ are observed.} 
\label{fig:figure5}
\end{figure}

The angular-dependent critical current densities ($J_{\rm c}(\Theta)$) at 8\,K in several magnetic fields, fig.\,\ref{fig:figure5}, always show a peak positioned at $\Theta=90^\circ$ and 270\,$^\circ$ owing to the intrinsic pinning.\cite{13} Here the magnetic field was applied in the maximum Lorentz force configuration ($H$\,$\perp$\,J) at an angle $\Theta$ measured from the $c$-axis. A small $J_{\rm c}$ anisotropy ($\gamma_{\rm J}=J_{\rm c}(90^\circ)/J_{\rm c}(180^\circ)$) of 2.6 is observed at 1\,T. This $\gamma_{\rm J}$ is significantly increasing with $H$ (e.g. $\gamma_{\rm J}=46.8$ at $\mu_0H=9$\,T) due to the relatively close to $H_{\rm irr}$ in the $c$ direction.  
  
In summary, epitaxial FeSe$_{1-x}$Te$_x$ films with sharp out-of-plane and in-plane texture have been realized on Fe-buffered single crystalline MgO substrates similar to the Fe/Ba-122 bilayers. These results indicate that Fe can work as generic buffer layer for epitaxial growth of Fe-based superconductors. The Fe/FeSe$_{1-x}$Te$_x$ bilayer with a high $T_{\rm c}$ of 17.7\,K showed strong intrinsic pinning from correlated $ab$-planes, since the Se(Te)-Se(Te) interlayer distance is almost identical to the out-of-plane coherence length at low temperatures. 

\begin{acknowledgments}
The authors thank J.\,Scheiter and E.\,Reich for help with FIB cut samples and TEM, and E.\,Barbara for help with the AFM observation. We are also grateful to M.\,K\"{u}hnel and U.\,Besold for their technical support and S.\,F\"{a}hler for his RHEED software. This work was partially supported by DFG under Project no. BE\,1749/13 and HA\,5934/3-1. We also acknowledge the EU (IRON-SEA and SUPERIRON) under Project no. FP7-283141 and FP7-283204. S.\,W. acknowledges support by DFG under the Emmy-Noether program (Grant no. WU595/3-1).
\end{acknowledgments}


%



\end{document}